\documentclass{appolb}
\usepackage{array}
\usepackage{color}
\usepackage{bm}
\usepackage{caption}
\usepackage{graphicx}
\newcommand{\ddbar}{$D\overline{D}$}
\newcommand{\dsdsbar}{$D_s\overline{D}_s$}
\newcommand{\ddabar}{$D\overline{D}^\ast$}
\newcommand{\dzdazbar}{$D^0\overline{D}^{\ast0}$}
\newcommand{\dpmdampbar}{$D^\pm\overline{D}^{\ast\mp}$}
\newcommand{\dsdsabar}{$D_s\overline{D}^\ast_s$}
\newcommand{\dadabar}{$D^\ast\overline{D}^\ast$}
\newcommand{\dsadsabar}{$D^\ast_s\overline{D}^\ast_s$}

\newcommand{\tpz}{${}^{3\!}P_0$}
\newcommand{\tpo}{${}^{3\!}P_1$}
\newcommand{\tpt}{${}^{3\!}P_2$}

\newcommand{\tft}{${}^{3\!}F_2$}

\newcommand{\spo}{${}^{1\!}P_1$}

\newcommand{\tspo}{$2\,{}^{1\!}P_1$}

\newcommand{\kap}{$K_0^\star(700)$}

\begin{document}
\title{Unquenched Radially Excited $P$-wave Charmonia\thanks
{Presented at ``Excited QCD 2026'', Granada, Spain, 9--13 Jan.\ 2026.}}
\author{
George Rupp\address{Centre for Theoretical Particle Physics, 
Instituto Superior T\'{e}cnico, \\
University of Lisbon, P-1049-001, Portugal}
}
\maketitle
\begin{abstract}
The ground-state positive-parity charmonia $\chi_{c0}(1P)$, $\chi_{c1}(1P)$,
$h_c(1P)$, and $\chi_{c2}(1P)$ are generally well described in
static (``quenched'') quark models, in which dynamical effects
of actual or virtual strong decay are neglected. In contrast, the five 
PDG candidates for $P$-wave charmonia in the energy region 3.85--3.95~GeV,
probably including the first radial excitations of the above ones, display a
totally different and quite disparate mass pattern. Moreover, two scalar
states are listed, viz.\ $\chi_{c0}(3860)$ and $\chi_{c0}(3915)$,
the former one apparently being very broad.

Preliminary results will be presented here for the first radial excitations of
the lowest $P$-wave $c\bar{c}$ states, obtained with the Resonance-Spectrum
Expansion while including in the calculation all OZI-allowed decay channels of
the most relevant charm-meson pairs. Employing a generalised scheme of
computing coupling constants for decays based on the ${}^{3\!}P_0$ model
ensures that no distortion of the spectra will occur due to the different
classes of allowed decay channels for the various positive-parity charmonia.
\end{abstract}
\section{Introduction}
Charmonium spectroscopy is undoubtedly the most fertile testing ground for
QCD-inspired quark models in view of the many observed states \cite{PDG24},
which allow to fine-tune model parameters and make further predictions, also
in the bottomonium sector. In particular, a good description of the lowest
positive-parity charmonia $\chi_{c0}(1P)$, $\chi_{c1}(1P)$, $h_c(1P)$, and
$\chi_{c2}(1P)$ with a spin-independent scalar linear-plus-Coulombic
confining potential is found, provided it is complemented with perturbative
spin-spin, spin-orbit, and tensor forces \cite{BCHLLW26,Burns14}.
%Using the experimental \cite{PDG24}
%values of the $1\,{^3P}_{0,1,2}$ states and Eqs.~(23--25) of Ref.~\cite{GI85},
%while setting for the moment the spin-spin contact term to zero as
%expected for $P$-wave charmonia, we obtain 34.95~MeV and 20.34~MeV for the
%spin-orbit and tensor contributions, respectively. (The values predicted in
%Ref.~\cite{GI85} were 28~MeV resp.\ 13~MeV). With these parameters, the
%$h_c(1P)$ mass is reproduced within 0.06~MeV, thus confirming the near
%vanishing of the $P$-wave contact term and also very strongly limiting the
%size of a possible vector component of the confining potential.
The status of the five PDG \cite{PDG24} candidates for radially excited
$P$-wave charmonia $\chi_{c0}(3860)$, $\chi_{c1}(3872)$, $\chi_{c0}(3915)$,
$\chi_{c2}(3930)$, and $X(3940)$ is completely different
%(see Table~\ref{two-P}).
(see Table~1).
\begin{table}[ht]
{
\small
\captionsetup{skip=2mm}
\caption{$2P$ charmonia candidates with listed \cite{PDG24} mass, width,
and decays.}
\begin{tabular}{|c|c|c|c|c|}
\hline
\rule[0pt]{0pt}{12pt}
PDG entry & $I^G(J^{PC})$ & M (MeV) & $\Gamma$ (MeV) & (Main) Decays \\
\hline\hline
\rule[-5pt]{0pt}{18pt}
\boldmath$\chi_{c0}(3860)$ & $0^+(0^{++})$ & $3862^{+26+40}_{-32-13}$ &
$201^{+154+88}_{-67-82}$ & $D^0\overline{D}^0 , \, D^+D^-$ \\
\hline
\rule[-5pt]{0pt}{18pt}
\boldmath$\chi_{c1}(3872)$ & $0^+(1^{++})$ & $3871.64 \pm 0.06$ &
$1.19 \pm 0.21$ & $D^0\overline{D}^0\pi^0 , \, \overline{D}^{\ast0}D^0$ \\
\hline
\rule[-5pt]{0pt}{18pt}
\boldmath$\chi_{c0}(3915)$ & $0^+(0^{++})$ & $3922.1 \pm 1.8$ &
$20 \pm 4$ & $D^+D^- , \, D^+_sD^-_s , \, \omega J/\psi$ \\
\hline
\rule[-5pt]{0pt}{18pt}
\boldmath$\chi_{c2}(3930)$ & $0^+(2^{++})$ & $3922.5 \pm 1.0$ &
$35.2 \pm 2.2$ & $D^+D^- , \, D^0\overline{D}^0$ \\
\hline
\rule[-5pt]{0pt}{18pt}
\boldmath$X(3940)$ & $?^?(?^{??})$ & $3942 \pm 9$ &
$43^{+28}_{-18}$ & $D^+D^- , \, D\overline{D}^\ast\,+\,$c.c. \\
\hline
\end{tabular}
}
\label{two-P}
\end{table}
For instance, two scalars instead of only one are listed, having wildly
disparate decay widths. Also,
$\chi_{c1}(3872)$ is surprisingly lighter than $\chi_{c0}(3915)$. Finally,
if $X(3940)$ is the \tspo\ state, as compatible with its observed \cite{PDG24}
$D\overline{D}^\star$ decay, being heavier than $\chi_{c2}(3930)$ is equally
unexpected. We can compare these irregularities to the situation in
bottomonium, where the corresponding $2P$ states all lie below the thresholds
of open-bottom meson pairs. Writing symbolically $\chi_{bi}(n)$ for
$M\!\left(\chi_{bi}(nP)\right)$ and $h_{b}(n)$ for
$M\!\left(h_{b}(nP)\right)$, we get \cite{PDG24} the following ratios of
$P$-wave mass splittings:
\[
\frac{\chi_{b1}(2)-\chi_{b0}(2)}{\chi_{b1}(1)-\chi_{b0}(1)}=0.69 \,, \;
\frac{h_{b}(2)-\chi_{b1}(2)}{h_{b}(1)-\chi_{b1}(1)}=0.67 \,, \;
\frac{\chi_{b2}(2)-h_{b}(2)}{\chi_{b2}(1)-h_{b}(1)}=0.69 \; .
\]
These numbers show a remarkably regular pattern of mass-splitting reductions
for the first $P$-wave $b\bar{b}$ radial excitations, which may very well be
the consequence of a node in the corresponding meson wave functions.
In contrast, the equivalent ratios among $P$-wave charmonia are
\[
\frac{\chi_{c1}(2)-\chi_{c0}(2)}{\chi_{c1}(1)-\chi_{c0}(1)}=-0.53\,, \;
\frac{h_{c}(2)-\chi_{c1}(2)}{h_{c}(1)-\chi_{c1}(1)}=4.79\, , \;
\frac{\chi_{c2}(2)-h_{c}(2)}{\chi_{c2}(1)-h_{c}(1)}=-0.63 \; .
\]
So clearly, the fact that the $2P$ charmonium states all lie \em above \em 
their lowest open-charm thresholds completely changes the picture, resulting
in complex mass shifts and threshold effects not governed by simple formulae.

Before investigating such phenomena explicitly for all $2P$
charmonia, I shall first review some old results for the $\chi_{c1}(3872)$
and $\chi_{c0}(3915)$ states.

\section{\boldmath$\chi_{c0}(3915)$ and \boldmath$\chi_{c1}(3872)$}
\begin{figure}[!t]
\centering
\includegraphics[trim = 0mm 0mm 0mm 0mm,clip,width=9cm,angle=0]
{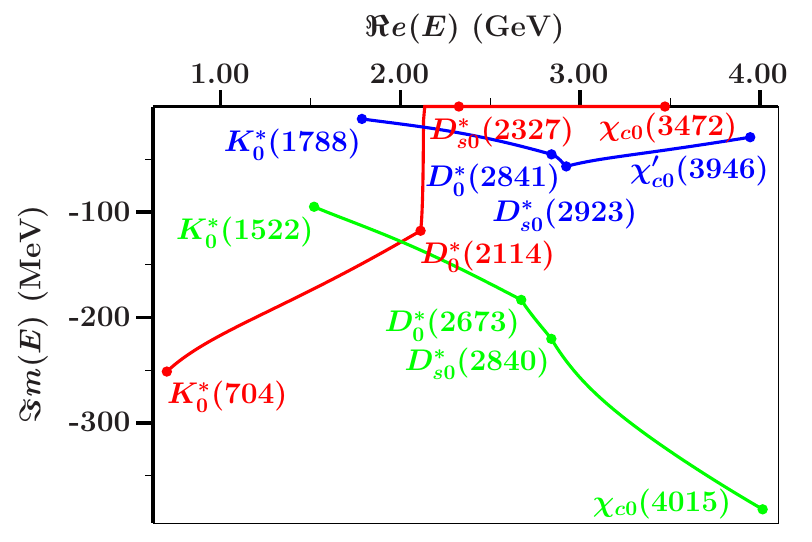}
\caption{Masses vs.\ widths of scalar mesons as a function of quark and
decay-product masses. Figure reprinted from Ref.~\cite{BCKR06} (arXiv
version).}
\label{masstrajectories}
\end{figure}

\begin{figure}[!t]
\centering
\includegraphics[trim = 0mm 0mm 0mm 0mm,clip,width=8cm,angle=0]
{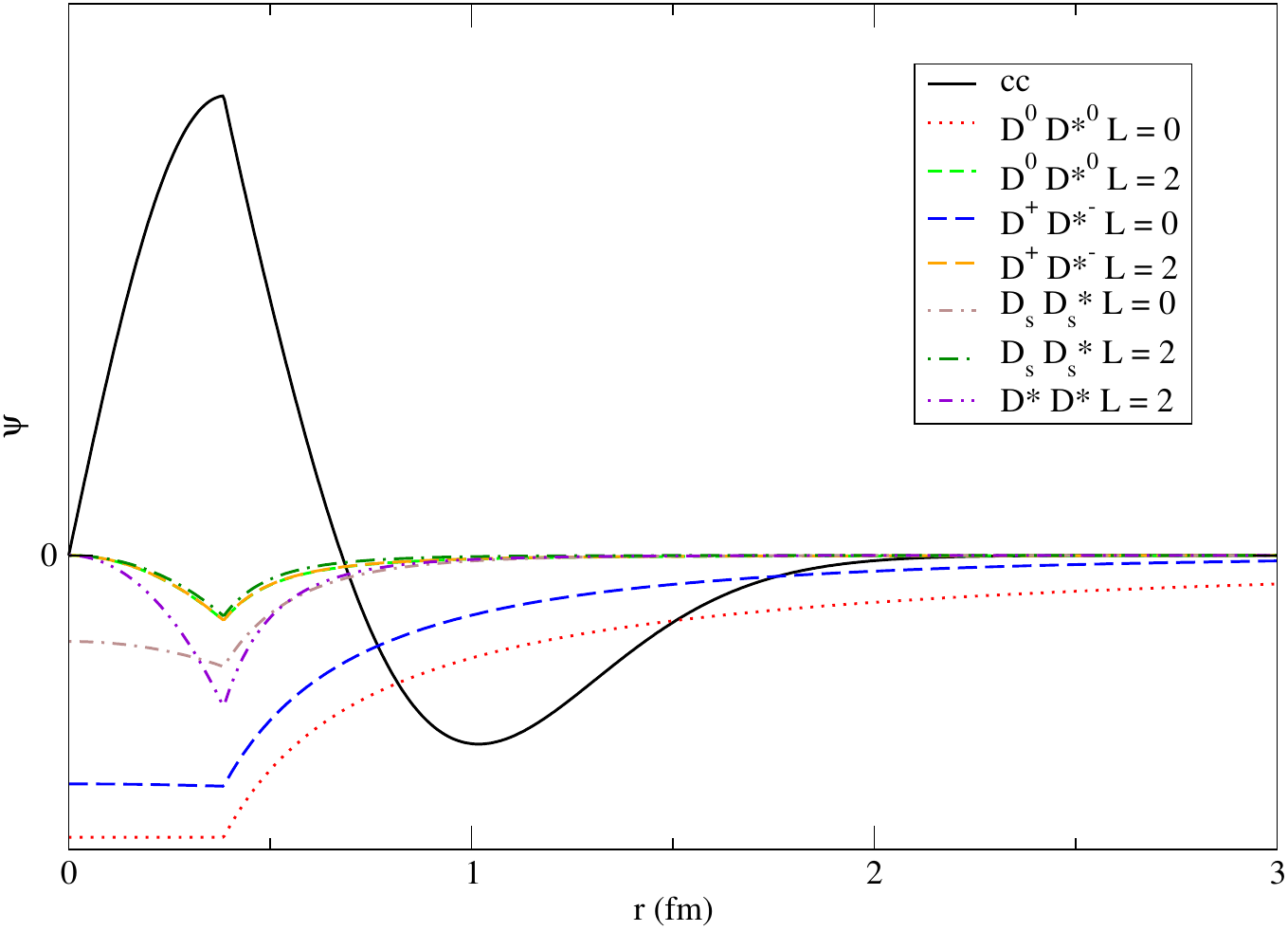}
\caption{$\chi_{c1}(3872)$ wave function in a multichannel model;
reprinted from Ref.~\cite{CRB15}.}
\label{wavefunction}
\end{figure}

In order to determine the nature of mesonic resonances in unitarised models,
pole trajectories as a function of coupling constant are often studied
(see e.g.\ some examples in Ref.~\cite{BR21} and the $\chi_{c1}(3872)$ case
below). However, much can be learned as well from trajectories as a function
of quark and decay-product masses. In Ref.~\cite{BCKR06}, a very simple model
of this type was employed to compute bound-state and resonance mass-width
trajectories of a dynamical, ground-state, and radially excited strange scalar
meson. This allowed to directly connect a variety of scalar mesons with one
another, including the \kap\ and the $\chi_{c0}(3915)$ \cite{PDG24}, the latter
resonance having first been observed by the Belle Collaboration in 2004
\cite{Belle05} at a mass of $3943\pm11$~MeV and found at 3946~MeV in
Ref.~\cite{BCKR06}. The corresponding trajectory is depicted in
Fig.~\ref{masstrajectories}, together with several others.
\begin{figure}[!b]
\centering
\includegraphics[trim = 30mm 170mm 55mm 9mm,clip,width=7cm,angle=0]
{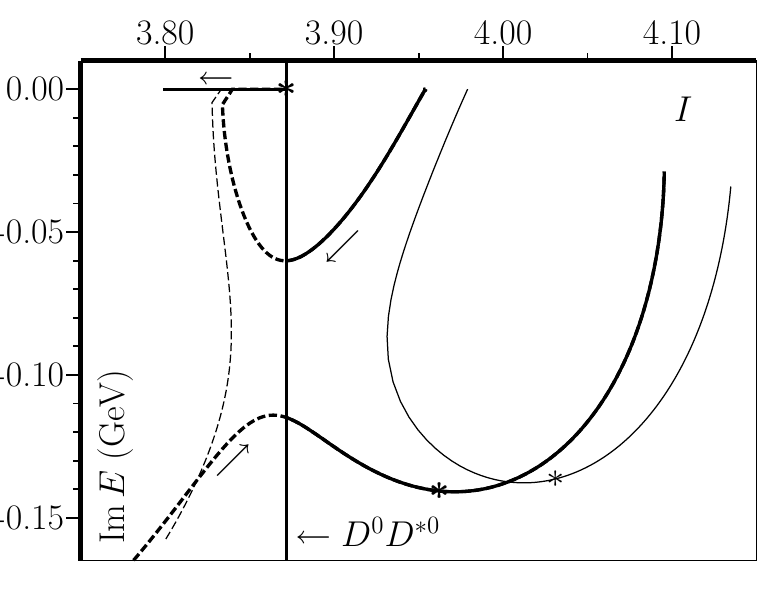}
\caption{
Pole trajectories as a function of decay coupling in a simple $r$-space
model. Boldface curves: $\chi_{c1}(3872)$ as a mass-shifted
intrinsic $P$-wave $c\bar{c}$ state; other curves: dynamical state.
Reprinted from Ref.~\cite{CRB13}.}
\label{poletrajectories}
\end{figure}
Concerning coupling-constant pole trajectories, in
Ref.~\cite{CRB13} the $\chi_{c1}(3872)$ meson was studied in a simple
unitarised model. Besides computing and plotting its two-component wave
function for varying parameters, it was also shown that a small
parameter change can make the state become a dynamical resonance instead
of an intrinsic one, as depicted in Fig.~\ref{poletrajectories}.
Finally, I recall the $\chi_{c1}(3872)$ wave function as obtained in a more
realistic unitarised model \cite{CRB15}, with different classes of decay
channels, as shown in Fig.~\ref{wavefunction}. One can see that at large
$r$ values the $\overline{D}^0D^{\ast0}$ component dominates, besides the
overall probability as well, but in the interior region the state is 
predominantly $c\bar{c}$.

\section{Charmonia with $J^{PC}=0^{++}$, $1^{++}$, $1^{+-}$, $2^{++}$ in the
RSE model}
Now I will proceed with the actual calculation of $2P$ charmonia employing
the Resonance-Spectrum Expansion (RSE) \cite{BR09}, very similarly to the
$\chi_{c1}(3872)$ modelling in Ref.~\cite{CRB11}, the main difference being
the additional inclusion here of the $D^\ast_s\overline{D}^\ast_s$ channel.
For the multichannel $T$-matrix in the RSE approach, see Appendix~A of 
Ref.~\cite{CRB11}. A crucial point in simultaneously studying resonances
with different quantum numbers is to ensure that no spectrum distortions will
occur owing to including different sets of decay channels. This can be
guaranteed by employing the formalism for computing decay couplings of
any ground-state or radially excited meson developed in Ref.~\cite{B83}.
The tables for \tpz, \tpo, \spo, and \tpt\ charmonia with the squares of their
couplings to the considered two-meson decay channels are presented in
Tables~\ref{tables12} and \ref{tables34}. Indeed, the sums of the squares of
\begin{table}[t!]
\small
\captionsetup{skip=2mm}
\caption{Decay couplings squared of $\chi_{c0}$ (left) and $\chi_{c1}$ (right)
states.}
\begin{tabular}{rl}
\begin{tabular}{|c|c|c|c|}
\hline
\rule[-8pt]{0pt}{24pt}
\boldmath$\chi_{c0}$ & $L$ & $\left[g_i^{(0)}\right]^2$ &
$\left[g_i^{(n)}\right]^2\!\times4^n$ \\ \hline\hline
\rule[-3pt]{0pt}{14.5pt} \ddbar     & 0 & 1/36  & $(n\!+\!1)/36$  \\
\hline
\rule[-3pt]{0pt}{14.5pt} \dsdsbar   & 0 & 1/72  & $(n\!+\!1)/72$  \\
\hline
\rule[-3pt]{0pt}{14.5pt} \dadabar   & 0 & 1/108 & $(n\!+\!1)/108$ \\
\hline
\rule[-3pt]{0pt}{14.5pt} \dsadsabar & 0 & 1/216 & $(n\!+\!1)/216$ \\
\hline
\rule[-3pt]{0pt}{14.5pt} \dadabar   & 2 & 5/27  & $(2n\!+\!5)/27$  \\
\hline
\rule[-3pt]{0pt}{14.5pt} \dsadsabar & 2 & 5/54  & $(2n\!+\!5)/54$  \\
\hline
\end{tabular}
&
\begin{tabular}{|c|c|c|c|}
\hline
\rule[-8pt]{0pt}{24pt}
\boldmath$\chi_{c1}$ & $L$ & $\left[g_i^{(0)}\right]^2$ &
$\left[g_i^{(n)}\right]^2\!\times4^n$ \\ \hline\hline
\rule[-3pt]{0pt}{16pt} \dzdazbar     & 0 & 1/54  & $(n\!+\!1)/54$  \\
\hline
\rule[-3pt]{0pt}{16pt} \dpmdampbar   & 0 & 1/54  & $(n\!+\!1)/54$  \\
\hline
\rule[-3pt]{0pt}{16pt} \dsdsabar     & 0 & 1/54  & $(n\!+\!1)/54$ \\
\hline
\rule[-3pt]{0pt}{16pt} \dzdazbar     & 2 & 5/216 & $(2n\!+\!5)/216$ \\
\hline
\rule[-3pt]{0pt}{16pt} \dpmdampbar   & 2 & 5/216 & $(2n\!+\!5)/216$ \\
\hline
\rule[-3pt]{0pt}{16pt} \dsdsabar     & 2 & 5/216 & $(2n\!+\!5)/216$ \\
\hline
\rule[-3pt]{0pt}{16pt} \dadabar      & 2 & 5/36  & $(2n\!+\!5)/36$ \\
\hline
\rule[-3pt]{0pt}{16pt} \dsadsabar    & 2 & 5/72  & $(2n\!+\!5)/72$ \\
\hline 
\end{tabular}
\end{tabular}
\label{tables12}
\end{table}
\begin{table}[b!]
\small
\captionsetup{skip=2mm}
\caption{Decay couplings squared of $h_{c}$ (left) and $\chi_{c2}$ (right)
states.}
\begin{tabular}{rl}
\begin{tabular}{|c|c|c|c|}
\hline
\rule[-8pt]{0pt}{24pt}
\boldmath$h_{c}$ & $L$ & $\left[g_i^{(0)}\right]^2$ &
$\left[g_i^{(n)}\right]^2\!\times4^n$ \\ \hline\hline
\rule[-3pt]{0pt}{14.5pt} \ddabar     & 0 & 1/54  & $(n+1)/54$ \\
\hline
\rule[-3pt]{0pt}{14.5pt} \dsdsabar   & 0 & 1/108 & $(n+1)/108$ \\
\hline
\rule[-3pt]{0pt}{14.5pt} \dadabar    & 0 & 1/54  & $(n+1)/54$ \\
\hline
\rule[-3pt]{0pt}{14.5pt} \dsdsabar   & 0 & 1/108 & $(n+1)/108$ \\
\hline
\rule[-3pt]{0pt}{14.5pt} \ddabar     & 2 & 5/54  & $(2n+5)/54$ \\
\hline
\rule[-3pt]{0pt}{14.5pt} \dsdsabar   & 2 & 5/108 & $(2n+5)/108$ \\
\hline
\rule[-3pt]{0pt}{14.5pt} \dadabar    & 2 & 5/54  & $(2n+5)/54$ \\
\hline
\rule[-3pt]{0pt}{14.5pt} \dsadsabar  & 2 & 5/108 & $(2n+5)/108$ \\
\hline
\end{tabular}
&
\begin{tabular}{|c|c|c|c|}
\hline
\rule[-8pt]{0pt}{24pt}
\boldmath$\chi_{c2}$ & $L$ & $\left[g_i^{(0)}\right]^2$ &
$\left[g_i^{(n)}\right]^2\!\times4^n$ \\ \hline\hline
\rule[-3pt]{0pt}{14.5pt} \dadabar    & 0 & 1/27  & $(n\!+\!1)/27$ \\
\hline
\rule[-3pt]{0pt}{14.5pt} \dsadsabar  & 0 & 1/54  & $(n\!+\!1)/54$ \\
\hline
\rule[-3pt]{0pt}{14.5pt} \ddbar      & 2 & 1/36  & $(2n\!+\!5)/180$ \\
\hline
\rule[-3pt]{0pt}{14.5pt} \dsdsbar    & 2 & 1/72  & $(2n\!+\!5)/360$ \\
\hline
\rule[-3pt]{0pt}{14.5pt} \ddabar     & 2 & 1/12  & $(2n\!+\!5)/60$ \\
\hline
\rule[-3pt]{0pt}{14.5pt} \dsdsabar   & 2 & 1/24  & $(2n\!+\!5)/120$ \\
\hline
\rule[-3pt]{0pt}{14.5pt} \dadabar    & 2 & 2/27  & $(n\!+\!1)/108+$ \\[-1mm]
\rule[-3pt]{0pt}{14.5pt} & & &                     $7(2n\!+\!5)/540$ \\
\hline
\rule[-3pt]{0pt}{14.5pt} \dsadsabar  & 2 & 1/27  & $(n\!+\!1)/216+$ \\[-1mm]
\rule[-3pt]{0pt}{14.5pt} & & &                     $7(2n\!+\!5)/1080$ \\
\hline
\end{tabular}
\end{tabular}
\label{tables34}
\end{table}
the ground-state $g_i^{(0)}$ are equal to 1/3 in all four cases.
The same equal sum of squares also holds for arbitrary $n$, except for 
$(n\!+\!1)$\tpt\ charmonia, which is probably related to mixing with the
$n$\tft\ states (see below).

The preliminary results of this calculation should not be taken at face value 
as for the precise numbers, in view of the omission so far of spin-orbit and
tensor splittings, but rather as an indication of unitarisation
effects for the different $2P$ $c\bar{c}$ states. Masses of the found
poles (in MeV): \\[-5mm]
\begin{eqnarray*}
\mbox{\tpz:}\;
\left\{\!\!\!\!
\begin{array}{cc}
\displaystyle
3871.4-i\times89.5 \\
3900.5-i\times36.5
\end{array}
\right. 
& , & \;\;\mbox{\tpo:} \;\, 3871.5-i\times0.7 \; , \\
\mbox{\spo:} \;\, 3877.0-i\times3.0
\hspace*{6.5mm} & , & \;\;\mbox{\tpt:} \;\, 3892.1-i\times0.3  \; .
\end{eqnarray*}
These numbers are obtained for the overall coupling $\lambda=3.1$ and
the decay radius $r=3.0$~GeV$^{-1}$, so equal or very close to the values
used in Ref.~\cite{CRB11}. Thus, the $\chi_{c1}(3872)$ pole comes out almost
on top of the average experimental result \cite{PDG24}, while the
$\chi_{c0}(3871.4-i\times89.5)$ pole is compatible with a $\chi_{c0}(3860)$
\cite{PDG24} having a large width $\sim\!200$~MeV, albeit with enormous error
bars. But more important than the precise model numbers is the fact that two
scalar $c\bar{c}$ resonances are found in this energy region. As for the
present $h_c(2P)$ and $\chi_{c2}(2P)$ results, it seems hard to explain the
$X(3940)$ \cite{PDG24} without also considering spin-orbit and tensor
splittings, besides possible \tpt/\tft\ mixing \cite{BR12}. These extensions
are already being studied in detail \cite{R26}, including a careful tracking
of poles in the complex energy and momentum planes in order to determine their
nature, as either dominantly intrinsic or dynamically generated 
positive-parity $c\bar{c}$ resonances.

\end{document}